\documentclass[12pt]{article}
\pdfoutput=1
\usepackage{putex}
\usepackage{graphicx}
\usepackage{epstopdf}
\usepackage{enumerate}
\usepackage{cite}
\usepackage{tensor}
\usepackage{slashed}

\numberwithin{equation}{section}

\newcommand{\HH}{\mathbb{H}}

\newcommand{\abs}[1]{\left\lvert #1 \right\rvert}

\newcommand {\be} {\begin {equation}}
\newcommand {\ee} {\end {equation}}

\newcommand {\bes} {\begin {equation*}}
\newcommand {\ees} {\end {equation*}}

\newcommand{\es}[2] {\begin{equation} \label{#1} \begin{split} #2 \end{split} \end{equation}}

\newcommand{\Z}{\mathbb{Z}}
\newcommand{\N}{\mathbb{N}}
\newcommand{\R}{\mathbb{R}}
\newcommand{\C}{\mathbb{C}}

\newcommand{\beq}{\begin{equation}}
\newcommand{\eeq}{\end{equation}}

\begin{document}

\institution{IAS}{School of Natural Sciences, Institute for Advanced Study, Princeton, NJ 08540}
\institution{MIT}{Center for Theoretical Physics, Massachusetts Institute of Technology, Cambridge, MA 02139}
\institution{Harvard}{Department of Physics, Harvard University, Cambridge, MA 02138}
\institution{PU}{Joseph Henry Laboratories, Princeton University, Princeton, NJ 08544}

\title{R\'enyi Entropies for Free Field Theories}

\preprint{PUPT-2398\\MIT-CTP-4329}

\authors{Igor R.~Klebanov,\worksat{\IAS}\footnote{On leave from
Joseph Henry Laboratories and Center for Theoretical Science, Princeton University.}
  Silviu S.~Pufu,\worksat{\MIT} Subir Sachdev,\worksat{\Harvard}
Benjamin R.~Safdi\worksat{\PU}}

\abstract{
R\'enyi entropies $S_q$ are useful measures of quantum entanglement; they can be calculated from traces of the reduced density matrix raised to power $q$, with $q \geq 0$.  For $(d+1)$-dimensional conformal field theories, the R\'enyi entropies across $S^{d-1}$ may be extracted from the thermal partition functions of these theories on either $(d+1)$-dimensional de Sitter space or $\R \times \HH^d$, where $\HH^d$ is the $d$-dimensional hyperbolic space.
These thermal partition functions can in turn be expressed as path integrals on branched coverings of the $(d+1)$-dimensional sphere and $S^1 \times \HH^d$, respectively.  We calculate the R\'enyi entropies of free massless scalars and fermions in $d=2$, and show how using zeta-function regularization one finds agreement between the calculations on the branched coverings of $S^3$ and on $S^1 \times \HH^2$.  Analogous calculations for massive free fields provide monotonic interpolating functions between the R\' enyi entropies at the Gaussian and the trivial fixed points.  Finally, we discuss similar R\'enyi entropy calculations in $d>2$.
}

\date{November 2011}

\maketitle

\section{Introduction}
\label{sec:intro}

The R\'enyi entropies \cite{renyi0,renyi1} have recently emerged as powerful diagnostics of long-range
entanglement in many-body quantum ground states in $d\geq 2$ spatial dimensions \cite{hastings,melko,pasquier,2011PhRvB..83x5134C,tarun1,qi1,swingle2,swingle1,2011arXiv1111.4836C,tarun2,square1,square2,maxgoldstone}.
The R\'enyi entropy $S_q$ is defined as
 \es{RenyiDef}{
  S_q = \frac{1}{1-q} \log \tr \rho^q \,, \qquad q \geq 0\,,
 }
where $\rho$ is the reduced density matrix obtained after tracing over the degrees of freedom in the complement of the entangling region.
Numerical evaluations of the entropies have allowed characterization of many distinct types
of ground states: gapped states with topological order \cite{melko,pasquier,2011PhRvB..83x5134C,tarun1,qi1},
quantum critical points \cite{swingle2}, Fermi liquids \cite{swingle1,2011arXiv1111.4836C},
non-Fermi liquids \cite{tarun2}, and Goldstone phases with broken symmetries \cite{square1,square2,maxgoldstone}. Indeed, it appears that
the R\'enyi entropies carry distinct signatures of the known quantum many-body states and
are also amenable to evaluation by convenient algorithms.

Nevertheless, there are few known analytic results for R\'enyi entropies with general $q$.
In $d=1$, corresponding to $D=d+1=2$ space-time dimensions, for
gapless conformal states, the Re\'nyi entropies of
an interval of length $R$ are given by \cite{cardy0,cardyCFT1,cardyCFT2}
 \es{Scalars2d}{
S_q = \frac{c}{6} \left( 1 + \frac{1}{q} \right) \log (R/\epsilon)\,,  \qquad d=1 \,,
 }
where $\epsilon$ is a short-distance cutoff, and $c$ is the central charge of
the conformal field theory (CFT)\@. For odd $d>1$, in studying R\'enyi entropies across the $(d-1)$-dimensional sphere of radius $R$,
one again finds terms logarithmic in $R$. Their coefficients as a function of $q$ were calculated in \cite{Casini:2010kt,Fursaev:1993hm,DeNardo:1996kp} for a massless real scalar field.
In the case $d=3$, the result is
\beq \label{fourds}
S_q^{\rm scalar} = \alpha \left( \frac{R}{\epsilon} \right)^2
-\frac{(1+q)(1+q^2)}{360 q^3}  \log (R/\epsilon)\,,
\eeq
where
$\alpha$ is a non-universal constant dependent upon ultraviolet details, and $\epsilon$ is again a short-distance cutoff.
For $q=1$ this result agrees with a direct numerical calculation of the entanglement entropy \cite{Lohmayer:2009sq}.
We note for completeness that logarithmic terms also appear in the R\'enyi entropy of a $d>1$
Fermi liquid \cite{swingle1,tarun2,2011arXiv1111.4836C},
$S_q \sim (1 + 1/q) R^{d-1} \ln R$, but these modify the leading `area law' term and so are much stronger than those
in
\eqref{fourds}.

In $d=2$, there is a simple known result for gapped topological states, such as those
described by an effective $U(1)$ Chern-Simons gauge theory at level $k$. The R\'enyi entropies
are {\em independent\/} of $q$ and given by \cite{tarun1,swingle2,hamma,levin,preskill}
\beq
S_q = \alpha { R \over \epsilon}  - \frac{1}{2} \ln k \,,
\label{sqcs}
\eeq
where $R$, $\alpha$, $\epsilon$ are as in (\ref{fourds}). Note that there is no logarithmic dependence on the size $R$
in the universal term. We reproduce this result in Appendix~\ref{cssection}.

Generally the structure of R\'enyi entropies in conformal field theories (CFTs) in $d=2$ is more complicated than the structure in $d=1$~\cite{kim,mfs,ch1}. In $d=2$ the leading term $\sim R/\epsilon$ is UV divergent and not universal.
However, the subleading $R$-independent term is finite and universal (in even $d$ such universal terms have no logarithmic dependence on $R$). Unlike in $d=1$, its dependence on $q$ is not simply through a general $q$-dependent normalization factor. More generally, we can deviate away from the CFT with a relevant operator
which generates a mass scale $m$; then the R\'enyi entropies obey \cite{mfs}
\beq
S_q = \alpha  \frac{R}{\epsilon}  - \mathcal{S}_q (m R)  \,,  \qquad d=2 \,,
\label{sqscale}
\eeq
where $\mathcal{S}_q (x)$ is a universal function, which in turn obeys
\beq
\mathcal{S}_q (x) = \begin{cases}
  r_q x  &  \text{as }x \rightarrow \infty \,, \\
  \gamma_q &  \text{as }x \rightarrow 0 \,,
\end{cases}
\eeq
where $r_q$ and $\gamma_q$ are universal numbers. The number $r_q$ describes the shift in the linear
$R$ dependence of $S_q$ due to the presence of the mass scale $m$, while $\gamma_q$
measures the universal contribution of the CFT to the R\'enyi entropy, similar to eq.~\eqref{sqcs}.
Ref.~\cite{mfs} obtained results for an infinite cylinder of circumference $R$
divided along a circular boundary: explicit results were obtained for all $q$ for both $\gamma_q$ and $r_q$
for the CFT of a free scalar field, while only $r_q$ was determined for general $q$ in the large $N$ limit
of the Wilson-Fisher CFT of $N$-component interacting scalars.

All our subsequent discussion of R\'enyi entropies will be restricted
to conformal field theories in which $m=0$. Also, we will not display
the leading non-universal term proportional to $R$, and so we write simply
\beq
S_q = - \gamma_q \,.
\eeq

In the limit $q\rightarrow 1$, the R\'enyi entropy (\ref{RenyiDef}) becomes the more familiar quantum entanglement entropy
(see \cite{cardyCFT1,Eisert:2008ur,Nishioka:2009un} for reviews and references to earlier work).
While in general its calculation is quite difficult, simplifications do occur for conformal field theories with
some particular geometries separating the entangling regions $A$ and $B$.  For example, in the case of entanglement across a sphere $S^{d-1}$ in $d$ flat spatial dimensions,
the use of conformal mappings reduces the calculation of entanglement entropy to that of the thermal entropy evaluated when the spatial geometry is taken to be the hyperbolic space $\HH^d$ \cite{Casini:2010kt,ch1,ch2}. For the purpose of calculating the entanglement entropy, the temperature is taken to be
$T_0=1/(2\pi R)$, where $R$ is the radius of curvature of $\HH^d$. For this special temperature, the required Euclidean path integral
calculation on $S^1 \times \HH^d$ is related by a further Weyl transformation to that on $S^{d+1}$ \cite{ch2}. For example, in $d=2$ the entanglement entropy
between a disk and its complement is given by $-F=\log |Z|$, where $Z$ is the Euclidean path integral on $S^3$. The latter quantity can be calculated in any 3-dimensional field theory with ${\cal N}\geq 2$ supersymmetry using the method of localization \cite{Kapustin:2009kz,Drukker:2010nc,Jafferis:2010un,Festuccia:2011ws}.
It has also been calculated in some simple non-supersymmetric CFTs, such as free field theories \cite{Casini:2010kt,ch1,Dowker:2010yj,Klebanov:2011gs}, the Wilson-Fisher fixed point of the $O(N)$ model for large $N$ \cite{Klebanov:2011gs}, and conformal gauge field theories with large numbers of flavors \cite{Klebanov:2011td}.

A separate approach to the computation of entanglement entropies \cite{rt} (see also \cite{Nishioka:2006gr,Klebanov:2007ws})
relies on the AdS/CFT correspondence \cite{Maldacena:1997re,Gubser:1998bc,Witten:1998qj}. This approach has been used to calculate entanglement entropy across $S^{d-1}$ in CFTs
that possess a dual description with weakly curved gravitational backgrounds \cite{sinha,ch2}.
All these results have led to conjectures that the 3-sphere free energy $F$, or equivalently minus the entanglement entropy across a circle,
decreases along any RG flow and is stationary at the fixed points \cite{sinha,Jafferis:2011zi,ch2,Klebanov:2011gs}.\footnote{After the original version of this paper appeared, a proof of the F-theorem was presented in \cite{Casini:2012ei}.}

The conformal methods for calculating the entanglement entropy $S_1$ have been extended recently \cite{headrick,Casini:2010kt,hung}
to allow computation of R\'enyi entropies $S_q$ in CFTs for spherical entangling geometries \cite{Casini:2010kt,hung}.
It turns out that now one needs the thermal free energy of the theory on $\HH^d$ evaluated at temperatures $T_0/q$.
It is useful to define
\be
{\cal F}_q = - \log \abs{ Z_q} \,,
\ee
where $Z_q$ is the Euclidean path integral on $S^1 \times \HH^{d}$, with $S^1$ of circumference $2\pi R q$. This geometry can be further
conformally mapped to a certain $q$-fold branched covering of $S^{d+1}$ analogous to the one described in the next section for $d = 2$. While this branched covering has curvature singularities, we will show that the path integral can often be simply calculated on this background, and the results agree with those on $S^1 \times \HH^{d}$.  The R\'enyi entropies can then be simply expressed as \cite{Casini:2010kt,hung}
\be
\label{basicform}
S_q= {q {\cal F}_1- {\cal F}_q\over 1- q}
\,.\ee
In \cite{hung} these methods were applied on the gravity side of the AdS/CFT duality, where the free energy is read off from that of a certain topological black hole geometry.
The goal of our paper is instead to continue along the lines of \cite{Casini:2010kt} and consider applications of these methods to free field theories, such as the conformally coupled scalar field and the massless Dirac fermion in $d=2$ ($D=3$).

We will present calculations both on $S^1 \times \HH^{2}$ and on the $q$-fold branched coverings of $S^{3}$ and demonstrate their consistency.  For simplicity we only consider integer $q$-fold branched coverings of $S^3$ (some further details may be found in Appendix A), though in principle the partition function on that space can be computed for any $q \geq 0$. We consider the conformal scalar field in section 3, and the massless Dirac fermion in section 4. In section 5 we show that analogous calculations for massive free fields provide monotonic interpolating functions between the R\' enyi entropies at the Gaussian and the trivial fixed points. In the Discussion section we comment on extensions to $d>2$.

\section{Methods for computing the R\'enyi entropy in CFT}
\label{RENYI}

In \cite{ch2,hung} it was explained how one can use conformal symmetry to express the R\'enyi entropy \eqref{RenyiDef} in terms of partition functions of the CFT on certain curved manifolds.  The curved manifolds relevant for $d=2$ are multiple branched coverings of the three-sphere
and $S^1 \times \HH^2$, where $\HH^2$ is the two-dimensional hyperbolic space.  One can see how these spaces arise by following the arguments presented in \cite{ch2, hung}. With the help of a Weyl transformation one can map the reduced density matrix on a disk of radius $R$ in $\R^{2, 1}$ to
 \es{GotRho}{
  \rho = \frac{e^{- 2 \pi R {\cal H}_t}}{ \tr e^{- 2 \pi R {\cal H}_t}} \,,
 }
where ${\cal H}_t$ is the Hamiltonian generating time translations in either: (A) the static patch of three-dimensional de Sitter space of radius $R$ with the metric
 \es{deSitter}{
  ds_A^2 =  - \cos^2 \theta dt^2 + R^2 \left[ d \theta^2 + \sin^2 \theta d \phi^2  \right] \,;
 }
or (B) $\R \times \HH^2$, with the metric
 \es{RHyp}{
  ds_B^2 = -dt^2 + R^2 \left[ d \eta^2 + \sinh^2 \eta\, d \phi^2 \right] \,.
 }
That the two spaces above differ just by a conformal rescaling can be seen by writing $ds_A^2 = \cos^2 \theta\, ds_B^2$, with the identification $\sinh \eta = \tan \theta$.

Using the definition of the R\'enyi entropies in eq.~\eqref{RenyiDef} and the density matrix \eqref{GotRho}, one finds
 \es{GotHs}{
  S_q = \frac{1}{1-q} \log \frac{\tr e^{- 2 \pi R q {\cal H}_t}}{(\tr e^{- 2 \pi R {\cal H}_t})^q}
   =  \frac{q{\cal F}_1 - {\cal F}_q}{1-q} =  \frac{2 \pi R q\left(F_1 - F_q \right)}{1-q}  \,,
 }
where we made the definitions
 \es{Partf}{
  Z_q = \tr e^{- 2\pi R q {\cal H}_t} = e^{-{\cal F}_q} \,, \qquad
   {\cal F}_q \equiv 2 \pi R q F_q\,,
 }
which hold for both of the mappings (A) and (B) introduced above.

The quantity $Z_q$ can be interpreted as the partition function at temperature $T = 1/(2 \pi R q)$ in either de Sitter space in the case \eqref{deSitter} or $\R \times \HH^2$ in the case \eqref{RHyp}, and $F_q = F(T)$ is the corresponding thermal free energy.  As is standard in thermal field theory, $Z_q$ can be computed as a Euclidean path integral after the Euclidean time direction has been compactified into a circle of length $\beta = 1/T = 2 \pi R q$.  When we use the mapping (A) to de Sitter space, the Euclidean metric is
 \es{Euclidean}{
  ds_A^2 = R^2 \left[ \cos^2 \theta d\tau^2 + d \theta^2 + \sin^2 \theta d \phi^2 \right] \,,
 }
where the ranges of the coordinates are $0 \leq \tau < 2 \pi q$, $0 \leq \phi < 2 \pi$, and $0\leq \theta < \pi/2$.  Similarly, when we use the mapping (B) to $\R \times \HH^2$, the Euclidean metric is
 \es{Metric}{
  ds_B^2 = R^2 \left[ d\tau^2 + d \eta^2 + \sinh^2 \eta\, d \phi^2 \right]\,,
 }
where the ranges of the coordinates are $0 \leq \tau < 2 \pi q$, $0 \leq \phi < 2 \pi$, and $0\leq \eta < \infty$.  Like their Minkowski counterparts \eqref{deSitter} and \eqref{RHyp}, these Euclidean spaces also differ from each other just by a conformal rescaling, as can be seen from writing $ds_A^2 = \cos^2 \theta\, ds_B^2$ with $\sinh \eta = \cos \theta$.

Before we examine the spaces \eqref{Euclidean} and \eqref{Metric} in more detail, let us comment on the case $q=1$, where eq.~\eqref{GotHs} seems poorly defined, and rewrite this equation in a different way.  Taking the limit $q\to 1$ in \eqref{GotHs}, one obtains
 \es{S1}{
  S_1 = \frac{dF(T)}{dT} \bigg|_{T=1/(2 \pi R)}
   = - S_\text{therm}(1/(2 \pi R)) \,,
 }
$S_\text{therm}(T)$ being the thermal free energy at temperature $T$.  In general, the definition of the free energy is
 \es{GeneralF}{
  F(T) = E(T) - T S_\text{therm}(T) \,,
 }
where $E(T) = \tr (\rho_q {\cal H}_t)$ is the total energy.  Precisely at $T = 1/(2 \pi R)$, or equivalently at $q=1$, the manifolds \eqref{Euclidean} and \eqref{Metric} are conformally equivalent to $\R^3$, which implies that for  either of these manifolds we must have $E(1/(2 \pi R)) = 0$.  It follows immediately from \eqref{S1} and \eqref{GeneralF} that $S_1 = -2 \pi R F_1 = -{\cal F}_1$.  A simple calculation then shows
 \es{Simple}{
  S_q - S_1 = \frac{{\cal F}_q - {\cal F}_1}{q-1}
   =2 \pi R \frac{q F_q - F_1}{q-1}  \,.
 }

Let us now comment on the features of the spaces \eqref{Euclidean} and \eqref{Metric}.  The space in eq.~\eqref{Euclidean} is not smooth unless $q = 1$, in which case it reduces to the round three-sphere of radius $R$.  One way to see this fact  is to think of $S^3$ as the sphere in $\C^2 \cong \R^4$ given by $\abs{z_1}^2 + \abs{z_2}^2 = R^2$.  We can write
 \es{ParamS3}{
  z_1 = R \cos \theta e^{i \tau} \,, \qquad
   z_2 = R \sin \theta e^{i \phi} \,,
 }
where $\theta$ ranges from $0$ to $\pi/2$, and $\tau$ and $\phi$ range from $0$ to $2 \pi$.  These coordinates are the same as those appearing in \eqref{Euclidean}, as one can easily check that when $q=1$ the standard line element in $\C^2$ takes the form \eqref{Euclidean} under the mapping given by \eqref{ParamS3}.  Making $\tau$ range from $0$ to $2 \pi q$ one then obtains \eqref{Euclidean} for arbitrary $q$.   The description of \eqref{Euclidean} in terms of the complex coordinates $z_1$ and $z_2$ is particularly useful when $q = 1/p$ for some positive integer $p$.  In this case the space \eqref{Euclidean} should be identified with the orbifold $S^3 / \Z_p$, the $\Z_p$ equivalence being given by $z_1 \sim z_1 e^{2 \pi i/p}$.  On the other hand, when $q$ takes values in the positive integers, \eqref{Euclidean} describes a $q$-fold covering of $S^3$ (henceforth denoted $C_q$) branched along the circle located at $\theta = \pi/2$.

One way of defining the spaces $C_q$, and really the way we will be thinking about them in this paper, is in terms of the normalizable functions (or more generally sections of spinor and vector bundles) that can be defined on them.  For instance, when $q=1$ a basis for the Hilbert space of normalizable functions consists of polynomials in $z_1$, $z_2$, and their complex conjugates;  when $q=2$, we are also allowed to have $\sqrt{z_1}$ or $\sqrt{z_1^*}$ times polynomials in $z_1$ and $z_2$, but not $\sqrt{z_2}$ and $\sqrt{z_2^*}$ times such polynomials, etc.  As we will see, the set of allowed functions and sections of spinor and vector bundles on $C_q$ is one of the central ingredients of our computations.

One of the advantages of working on the spaces \eqref{Euclidean} as opposed to \eqref{Metric} is that these spaces are compact, and after removing the UV divergences (for example by zeta-function regularization) the free energies $F_q$ are finite.  On the other hand, the hyperbolic cylinder \eqref{Metric} (henceforth denoted $H_q$) is non-compact, and regularization of UV divergencies yields a finite free energy density.  The free energy itself is proportional to the volume of $\HH^2$, which is infinite and requires further regularization.  The proper regularization of this volume \cite{Casini:2011kv} uses a hard cutoff at some value $\eta = \eta_0$:
 \es{VolH2}{
  \Vol(\HH^2) = 2 \pi \int_0^{\eta_0} d\eta\, \sinh \eta = 2 \pi \left[\frac {e^{\eta_0}}{2} - 1 + \frac{e^{-{\eta_0}}}{2} \right] \,.
 }
Taking the finite part of this expression, we should identify the regularized volume as
 \es{RegVolume}{
  \Vol(\HH^2) = -2 \pi \,.
 }

\section{R\'enyi entropies for free conformal scalars}
\label{BOSONSS3}

Among the simplest conformally invariant field theories in $d+1= D>2$ are free scalar fields $\phi$ conformally coupled to curvature.  The action for such a complex field on a manifold $M$ is
  \es{Action}{
  S = \int d^D r \sqrt{g(r)} \left[ \abs{\partial_\mu \phi}^2 + \frac {D-2}{4(D-1)}{\cal R} \abs{\phi}^2  \right] \,,
 }
where ${\cal R}$ is the Ricci scalar of $M$.  In this section we will set $D=3$ and work on the spaces $C_q$ and $H_q$ introduced in the previous section.  These spaces have constant scalar curvature ${\cal R}$, so the path integral yields
 \es{FreeConformal}{
  {\cal F} = \tr \log \left(-\Delta + \frac{\cal R}{8} \right) \,,
 }
where $\Delta$ is the Laplace operator.  We will first calculate ${\cal F}$ when $M = C_q$ (${\cal R} = 6$) and then reproduce the same results by taking $M = H_q$ (${\cal R} = -2$) (from here on we set the radius $R=1$).

In general, ${\cal R}=-d(d-1)$ on $S^1\times \HH^d$. Thus, the conformally coupled scalar (\ref{Action}) has
\es{GenConf}{
M^2=\frac {d-1}{4 d}{\cal R}=-\frac{(d-1)^2}{4}\,.
}
This means that the constant mode on $S^1$ saturates the Breitenlohner-Freedman (BF) bound \cite{Breitenlohner:1982jf}
on $\HH^d$.

\subsection{Conformal scalars on the $q$-fold branched covering of $S^3$}

The computation of ${\cal F}$ on $C_q$ starts with the diagonalization of the Laplacian on this space (in this section we take $q$ to be a positive integer).   We will show that the Laplacian has eigenvalues
 \es{LapEvalues}{
  \lambda_n = -n(n+2)
 }
for every $n \in \N / q$,\footnote{We use $\N$ to denote the set of non-negative integers (including zero).} with degeneracy
 \es{DegBosons}{
  g_n = \begin{cases}
   (n+1)^2\,, & \text{if $n \in \N$} \,,\\
   \left(k+1 \right) \left(k+2\right)\,, & \text{if }n=k + \frac pq\,, \qquad k, p \in \N \,, \qquad 1 \leq p < q \,.
  \end{cases}
 }

This result can be proven by examining the differential equation satisfied by an eigenfunction of the Laplacian.  The space $C_q$ has two $U(1)$ isometries corresponding to shifts in $\tau$ and $\phi$, so the eigenfunctions of the Laplacian can be assumed to be of the form $f(\theta) e^{i m_\tau \tau + i m_\phi \phi}$ for some $m_\tau$, $m_\phi$, and $f(\theta)$.  The eigenfunction equation for $f$ with eigenvalue $\lambda$ is
 \es{Evaluef}{
  f''(\theta) + 2 \cot \theta f'(\theta) - \left(\frac{m_\tau^2}{\cos^2\theta} + \frac{m_\phi^2}{\sin^2\theta} \right) f(\theta) = \lambda f(\theta) \,.
 }
The solution regular at $\theta = \pi/2$ is
 \es{fRegPiHalf}{
  f(\theta) &= (\cos \theta)^{\abs{m_\tau}} (\sin \theta)^{m_\phi} \\
    &\times {}_2F_1 \left(\frac{1 + \abs{m_\tau} + m_\phi - \sqrt{1- \lambda}}{2},  \frac{1 + \abs{m_\tau} + m_\phi + \sqrt{1- \lambda}}{2},
    1 + \abs{m_\tau}, \cos^2\theta \right) \,.
 }
Regularity at $\theta = 0$ implies
 \es{Gotlambdascalar}{
  \lambda = -n(n+2) \,, \qquad n = \abs{m_\tau} + \abs{m_\phi} + 2 a\,, \qquad  a \in \N \,.
 }
Note that $n$ itself is not required to be an integer.   As discussed towards the end of section~\ref{RENYI}, the allowed values of $m_\tau$ and $m_\phi$ are related to the periodicities of $\tau$ and $\phi$, respectively:  the period of $\phi$ is $2 \pi$ and that requires $m_\phi \in \Z$;  the period of $\tau$ is $2 \pi q$, and the condition for having well-defined eigenfunctions is $m_\tau \in \Z/q$.  It follows that $n \in \N / q$, as claimed above.  It is then straightforward to show by induction that the number of ways one can choose $m_\tau$, $m_\phi$, and $a$ so that $n =   \abs{m_\tau} + \abs{m_\phi} + 2 a$ is given by \eqref{DegBosons}.

According to \eqref{FreeConformal} the free energy is
 \es{FreeBosons}{
  {\cal F}_q = \sum_{k=1}^\infty k^2 \log \left[ k^2 - \frac 14 \right]
   + \sum_{p=1}^{q-1} \sum_{k=1}^\infty k(k+1) \log\left[\left(k + \frac pq\right)^2
   - \frac 14 \right] \,.
 }
This expression is divergent and requires regularization.  The first step in obtaining a regularized expression is to write $\log\left[\left(k + \frac pq\right)^2  - \frac 14 \right]  = \log\left[k + \frac pq+ \frac 12 \right]  + \log\left[k + \frac pq  - \frac 12 \right] $, and after a rearrangement of the terms in the sums one obtains
 \es{FreeBosonsAgain}{
  {\cal F}_q =  \sum_{k=1}^\infty (2k^2 - 2k + 1) \log \left[k - \frac 12 \right] + 2 \sum_{p=1}^{q-1} \sum_{k=1}^\infty k^2 \log\left[k + \frac pq
   - \frac 12 \right] \,.
 }
We can now use zeta-function regularization to write
 \es{FreeBosonsDer}{
  {\cal F}_q =  -\frac {d}{ds} \left[ \sum_{k=1}^\infty \frac{2k^2 - 2k + 1}{\left( k - \frac 12 \right)^s} + 2 \sum_{p=1}^{q-1} \sum_{k=1}^\infty \frac{k^2}{\left(k + \frac pq - \frac 12 \right)^s} \right] \Biggr\rvert_{s=0} \,.
 }
The sums appearing in \eqref{FreeBosonsDer} are easily evaluated in terms of Hurwitz zeta-functions:
 \es{FreeBosonsHurwitz}{
  {\cal F}_q &=  -2\left[ \zeta'(-2, 1/2) + \frac 14 \zeta'(0, 1/2)  \right] - 2 \sum_{p=1}^{q-1} \Biggl[ \zeta'\left(-2, \frac pq + \frac 12 \right) \\ &\qquad\qquad {}- 2 \left(\frac pq - \frac 12 \right) \zeta'\left(-1, \frac pq + \frac 12 \right) + \left(\frac pq - \frac 12 \right)^2 \zeta'\left(0, \frac pq + \frac 12 \right) \Biggr] \,,
 }
where the derivative acts on the first argument of the Hurwitz zeta-function.

It is also useful to write ${\cal F}_q$ in a more elementary form using the identities \eqref{DerZetaIdentity} presented in Appendix~\ref{MATH}.  With the help of these identities one can show
 \es{FreeBosonsSums}{
  {\cal F}_q = \frac{q}{2 \pi^2} \sum_{\substack{n=1\\ q \mid n}}^\infty \frac{(-1)^{n}}{n^3}
   - \frac{1}{2 \pi} \sum_{\substack{n=1\\ q\nmid n}}^\infty \frac{\cot (n \pi / q) (-1)^n}{n^2}
    - \frac 1{2q}  \sum_{n=1}^\infty \frac{a_n}{n} \,,
 }
 where
  \es{Gotan}{
   a_n = \begin{cases}
    (-1)^n \csc^2 (n \pi / q) & \text{if $q \nmid n$}\,, \\
    (-1)^n \frac{q^2 + 2}{6}  & \text{if $q \mid n$} \,.
   \end{cases}
  }
Note that all the sums in eq.~\eqref{FreeBosonsSums} are convergent:  the first two are absolutely convergent and the third converges because $\sum_{n=m q + 1}^{(m+2) q} a_n = 0$ for any $m \in \N$.

The first few free energies are
 \es{FreeBosonExamples}{
  {\cal F}_1 &= \frac{\log 2}{4} - \frac{3 \zeta(3)}{8 \pi^2} \,, \\
  {\cal F}_2 &=  \frac{\log 2}{4} + \frac{\zeta(3)}{8 \pi^2}  \,, \\
  {\cal F}_3 &=  \frac{\log 2}{4} - \frac{\zeta(3)}{24 \pi^2}
   + \frac{\psi_1(1/6) + \psi_1(1/3) - \psi_1(2/3) - \psi_1(5/6)}{72 \sqrt{3} \pi} \,, \\
  {\cal F}_4 &= \frac{5 \log 2}{16} + \frac{\zeta(3)}{32 \pi^2} + \frac{G}{2 \pi} \,,
 }
where $\psi_1(x)$ is the digamma function, and $G$ is the Catalan constant.  It is also not hard to find the large $q$ asytmptotics of ${\cal F}_q$.  When $q$ is large, we can approximate $\cot (n \pi / q) \approx \csc(n \pi /q) \approx q / (n \pi)$ in \eqref{FreeBosonsSums}, and so
 \es{FInfBosons}{
  {\cal F}_q \approx - \frac{q}{ \pi^2} \sum_{n=1}^\infty \frac{(-1)^n}{n^3} = \frac{3 \zeta(3)}{4 \pi^2} q \qquad
   \text{at large $q$}\,.
 }

\subsection{Conformal scalar on the hyperbolic cylinder} \label{scalarsHyp}

 In this section we calculate ${\cal F}_q$ on the hyperbolic cylinder $H_q$ introduced in eq.~\eqref{Metric} (now we do not restrict $q$ to be an integer). In anticipation of section~\ref{MASSIVE} below, we start by considering a massive complex scalar field on $H_q$ with the action
 \es{BosonAction}{
  S = \int d^3 r \sqrt{g(r)} \left[\abs{\partial_\mu \phi}^2 + M^2 \abs{\phi}^2 \right] \,.
 }
The conformally coupled scalar \eqref{Action} corresponds to $M^2 = -1/4$, and we will shortly specialize to this value of $M^2$.  It is convenient to define $m^2 = M^2+\frac 1 4$, so that the conformally coupled scalar has $m^2 = 0$.  Since on the $\HH^2$ of unit radius the BF bound \cite{Breitenlohner:1982jf} is $M^2\geq M_{BF}^2=-\frac 1 4$, we note that  $m^2 = M^2- M^2_{BF}$.

The free energy on $H_q$ is
 \es{FreeMassive}{
  {\cal F}_q(m^2) = \tr \log \left(-\Delta - \frac 14 +  m^2 \right) \,.
 }
The operator under the log is diagonalized by wavefunctions of the form $f(\eta, \phi) e^{i n_\tau \tau/q}$ for integer $n_\tau$ and for $f$ being an eigenfunction of the Laplacian on $\HH^2$.  Since $\HH^2$ is not compact, the spectrum of the Laplacian on it is continuous, consisting of eigenvalues $\lambda + 1/4$ for $\lambda \geq 0$ with the density of states
 \es{DLambda}{
  {\cal D}(\lambda) d\lambda = \frac{\Vol(\HH^2)}{4 \pi} \tanh (\pi \sqrt{\lambda}) d\lambda \,.
 }
(see for example \cite{camp2, MR1178146, Bytsenko:1994bc}).  The eigenvalues of $-\Delta  + m^2 - 1/4$ on $H_q$ form towers of continuous states for each $n_\tau$, each tower consisting of eigenvalues
 \es{Evalues}{
  \lambda + \frac {n_\tau^2}{q^2} + m^2 \,,
 }
with the same density of states as \eqref{DLambda}.   The free energy \eqref{FreeMassive} becomes
 \es{FreeScalarHyp}{
  {\cal F}_q(m) &= \sum_{n_\tau = -\infty}^\infty  \int_0^\infty d \lambda {\cal D}(\lambda) \log \left(\lambda  + \frac {n_\tau^2}{q^2} + m^2 \right) \,.
 }

We can use \eqref{BosonFermionParticular} to compute the regularized sum over $n$ and we obtain
 \es{FqOneInt}{
  {\cal F}_q(m) = \int_0^\infty d \lambda {\cal D}(\lambda)
   \left[ 2 \log \left(1 - e^{-2 \pi q \sqrt{\lambda + m^2 } }  \right)
    + 2 \pi q  \sqrt{\lambda + m^2 }\right] \,.
 }
We will use this formula for arbitrary $m \geq 0$ in section~\ref{MASSIVE}, but for now we set $m = 0$.
The integral in eq.~\eqref{FqOneInt} is divergent, but the divergence is in the temperature-independent vacuum energy given by the second term in \eqref{FqOneInt}.  This divergence can be regulated with zeta-function regularization or equivalently by subtracting the  $\R^3$ free energy density
 \es{FlatFree}{
 q \frac{\Vol(\HH^2)}{2} \int d\lambda \sqrt{\lambda} \,.
 }
Performing the integral of the difference between the second term in \eqref{FqOneInt} and \eqref{FlatFree}, and using $\Vol(\HH^2) = - 2 \pi$ as explained at the end of section~\ref{RENYI}, we obtain
 \es{ConformalAnswer}{
  {\cal F}_q \equiv {\cal F}_q(0) = - \int_0^\infty d \lambda \tanh(\pi \sqrt{\lambda})
    \log \left(1 - e^{-2 \pi q \sqrt{\lambda} }  \right)
     + q \frac{3 \zeta(3)}{4\pi^2} \,.
 }
This expression is convergent for any $q>0$ and it represents one of our main results.

While we are not aware of a general formula for the integral in \eqref{ConformalAnswer}, when $q$ is an integer the expression \eqref{ConformalAnswer} evaluates to the free energies \eqref{FreeBosonsHurwitz}---\eqref{FreeBosonExamples} we found in the previous subsection from the $q$-fold branched covering of $S^3$.  For example, when $q=1$, one can evaluate the first integral:
 \es{FirstIntegral}{
   - \int_0^\infty d \lambda \tanh(\pi \sqrt{\lambda})
    \log \left(1 - e^{-2 \pi \sqrt{\lambda} }  \right) =  \frac{\log 2}{4} - \frac{9 \zeta(3)}{8 \pi^2} \,,
 }
which, when combined with \eqref{ConformalAnswer} yields precise agreement with \eqref{FreeBosonExamples}:
 \es{GotF1}{
  {\cal F}_1 = \frac{\log 2}{4} - \frac{3 \zeta(3)}{8 \pi^2} \,.
 }
Similarly,
 \es{F2MinusF1}{
   {\cal F}_2 - {\cal F}_1  =\int_0^\infty d\lambda \tanh(\pi \sqrt{\lambda}) \log \frac{1 + \tanh(\pi \sqrt{\lambda})}{2} + \frac{3 \zeta(3)}{8 \pi^2}  = \frac{\zeta(3)}{2 \pi^2} \,,
 }
also in agreement with \eqref{FreeBosonExamples}, etc. In the limit of large $q$, the second term in (\ref {ConformalAnswer}) clearly dominates, and its value agrees with (\ref{FInfBosons}).

As another consistency check, we can demonstrate by explicit computation that the total energy at temperature $T_0 = 1/(2 \pi)$ vanishes.  Indeed, in general
 \es{ET0}{
  E = F - T \frac{dF}{dT} = \frac{1}{2 \pi }\frac{d{\cal F}_q}{dq} \,.
 }
Differentiating our general formula \eqref{ConformalAnswer} with respect to $q$, we obtain
 \es{F1Prime}{
  E(T_0) = \frac{1}{2 \pi } \frac{d{\cal F}_q} {dq} \Biggr \rvert_{q=1} = - \int_0^\infty d\lambda \frac{\sqrt{\lambda}}{1 + e^{2 \pi \sqrt{\lambda}}} +  \frac{3 \zeta(3)}{8\pi^3} = 0 \,,
 }
confirming the conceptual argument used in section~\ref{RENYI} that yielded the same result.

\subsection{Results for the R\'enyi entropies of a complex conformal scalar}

\begin{figure}[htb]
\begin{center}
\leavevmode
\scalebox{1}{\includegraphics{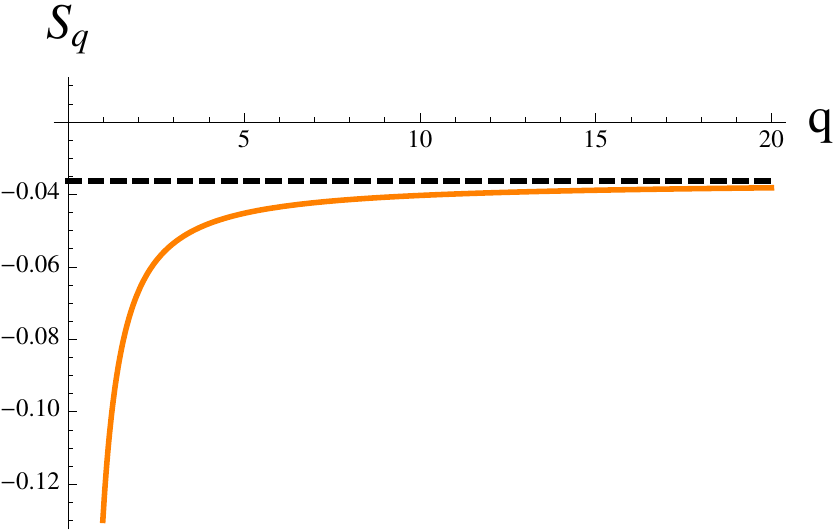}}
\end{center}
\caption{ The R\'enyi entropies $S_q$ for the complex conformal scalar field.  Note that the function $S_q$ is a monotonic function of $q$.  The black dashed line is the asymptotic value $S_\infty$. }
\label{Sqs}
\end{figure}

Now that we have checked the agreement between the calculations of ${\cal F}_q =- \log \abs{ Z_q}$ on $C_q$ and $H_q$, we can use (\ref{basicform}) to calculate the R\'enyi entropies. Some of the results for integer $q$ are listed below:
 \es{SummaryBosons}{
  S_1 &= -{\cal F}_1=  - \frac{\log 2}{4} + \frac{3 \zeta(3)}{8 \pi^2} \approx -0.1276 \\
  S_2 &= {\cal F}_2- 2 {\cal F}_1 =- \frac{\log 2}{4} + \frac{7 \zeta(3)}{8 \pi^2} \approx -0.0667 \\
  S_3 &= \frac 1 2  {\cal F}_3 - \frac 3 2 {\cal F}_1= - \frac{\log 2}{4} + \frac{13 \zeta(3)}{24 \pi^2}
  +\frac{\psi_1(1/6) + \psi_1(1/3) - \psi_1(2/3) - \psi_1(5/6)}{144 \sqrt{3} \pi} \approx -0.0534  \\
  S_4 &= \frac 1 3  {\cal F}_4 - \frac 4 3  {\cal F}_1= - \frac{11\log 2}{48} + \frac{49 \zeta(3)}{96 \pi^2} +\frac {G} {6\pi}
  \approx -0.0481\\
  S_\infty &=-{\cal F}_1+\lim_{q\rightarrow \infty} \frac {{\cal F}_q} {q}  = - \frac{\log 2}{4} + \frac{9 \zeta(3)}{8 \pi^2} \approx -0.0363 \,.
 }
 At large $q$ we find that
\es{sqlarge}{
S_q = S_\infty \left( 1 + \frac 1 q + O(1/q^2) \right) \, .
}
A plot of $S_q$ is presented in Fig.~\ref{Sqs}.
We note that, as in the holographic calculations of \cite{hung}, $S_q$ is a monotonically increasing function of $q$.
It would be interesting to compare (\ref{SummaryBosons}) with direct numerical calculations similar to those performed for $d=3$ in \cite{Lohmayer:2009sq}.

\section{R\'enyi entropies for free fermions}

Another simple conformal field theory is that of free fermions.  On a manifold $M$ of dimension $D$, the action is
 \es{ActionFermions}{
  S = \int d^D r \, \sqrt{g(r)}\, \bar \psi (i \slashed{D}) \psi \,,
 }
where $i \slashed{D}$ is the Dirac operator.  The factor of $i$ appears in the action such that the operator $i \slashed{D}$ is Hermitian, and hence it has real spectrum.  The free energy one obtains from integrating out $\psi$ is
 \es{FreeFermionsDef}{
  \tilde {\cal F} = - \tr \log (i \slashed{D}) \,.
 }
We will now compute $ \tilde {\cal F}$ in the case $M = C_q$ and then in the case $M = H_q$, in $D = 3$.  We then use eq.~\eqref{GotHs} to extract the R\'enyi entropies.  We will only study Dirac fermions, which are two-component complex spinors in $D=3$.

\subsection{Free fermions on the $q$-fold branched covering of $S^3$}
\label{FERMIONSS3}

In order to compute $\tilde {\cal F}$ on $C_q$, one first has to diagonalize the Dirac operator.  The diagonalization can be done by examining the eigenvalue problem in differential form, as was done for the Laplace operator in section~\ref{BOSONSS3}.  The analysis in the case of the Dirac operator is more involved that that in section~\ref{BOSONSS3}, because one now has to deal with spinors as opposed to functions, and we leave the details for Appendix~\ref{DEGFERMIONS}.   Here let us just quote the result:  the eigenvalues of the Dirac operator on the $q$th branched covering $C_q$ of $S^3$ form $q$ towers parameterized by an integer $p$ with $0 \leq p < q$.  For each $p$ the eigenvalues are
 \es{DiracEvalues}{
   \pm \left(k + 1 + \frac pq + \frac 1{2q} \right) \,, \qquad k \in \N \,,
 }
with degeneracy
 \es{DegFermions}{
   (k+1)(k+2)
 }
for each choice of sign.   For example, when $q=1$ we have only one tower of eigenvalues $k + 3/2$;  when $q=2$ we have two towers, one consisting of $k + 3/2 - 1/4$ and one of $k + 3/2 + 1/4$, etc.  The various towers are symmetric about $k + 3/2$, but the numbers $k + 3/2$ are part of the spectrum only for odd $q$.

The free energy following from the eigenvalues \eqref{DiracEvalues} and degeneracies \eqref{DegFermions} is
 \es{FqFermions}{
  \tilde {\cal F}_q =  - 2 \sum_{p=0}^{q-1} \sum_{k=0}^\infty k(k+1) \log \left(k+\frac pq + \frac 1{2q} \right) \,.
   }
This expression is of course divergent, but after zeta-function regularization it can be put in the form
 \es{FqFermionsZeta}{
  \tilde {\cal F}_q = 2 \sum_{p=0}^{q-1} \Biggl[\zeta'\left(-2, \frac pq + \frac 1{2q} \right) -2 \left( -\frac 12 + \frac pq + \frac 1{2q} \right) \zeta'\left(-1, \frac pq + \frac 1{2q} \right) \\
  +\left(\left( -\frac 12 + \frac pq + \frac 1{2q} \right)^2 - \frac 14 \right) \zeta'\left(0, \frac pq + \frac 1{2q} \right) \Biggr] \,,
 }
where the derivatives act on the first argument of the Hurwitz zeta-function as before.  While this is our final answer for the free energy of fermions on $C_q$, it may be instructive to put it in a form similar to eq.~\eqref{FreeBosonsSums} from the boson case by using the zeta-function identities \eqref{DerZetaIdentity}:
 \es{FreeFermionsText}{
  \tilde {\cal F}_q &= \frac{3 \zeta(3)}{8 \pi^2 q^2} + \frac{1}{2 \pi} \sum_{\substack{n=1\\ q \nmid n}}^\infty \frac{\csc (n \pi / q)}{n^2}
   +  \frac{1}{2 q} \sum_{\substack{n=1\\ q \nmid n}}^\infty \frac{\cos (n \pi /q) }{n \sin^2 (n \pi/q)}
     + \frac{2q^2 + 1}{12 q^2}\log 2  \,.
 }
These sums can be evaluated in terms of polygamma functions, but the resulting expression is not very enlightening.

The first few $\tilde {\cal F}_q$ can be obtained from evaluating the sums:
 \es{FreeFermionsExamples}{
  \tilde {\cal F}_1 &=   \frac{3 \zeta(3)}{8 \pi^2} + \frac{\log 2}{4} \,, \\
  \tilde {\cal F}_2 &= \frac{3 \zeta(3)}{32 \pi^2} + \frac {3 \log 2}{16} + \frac{G}{2 \pi} \,, \\
  \tilde {\cal F}_3 &=  \frac{\zeta(3)}{24 \pi^2} + \frac{\psi_1(1/3)}{3 \sqrt{3} \pi} + \frac{\log 2}{4} - \frac{2 \pi}{9 \sqrt{3}} \,.
 }
As in the case of bosons, it is also possible to find the large $q$ asymptotics of $\tilde {\cal F}_q$ by making a small angle approximation inside the trigonometric functions:
 \es{FreeLargeq}{
  \tilde {\cal F}_q \approx  \frac{q}{\pi^2} \sum_{n=1}^\infty \frac{1}{n^3}
   = \frac{\zeta(3)}{\pi^2} q \qquad \text{at large $q$}\,.
 }

\subsection{Fermions on the hyperbolic cylinder}  \label{fermionsHyp}

Like for bosons, we can reproduce the results of the previous subsection from a computation on the space $H_q$.  Let us start with the action for a massive Dirac fermion of mass $m$ and set $m =0$ later on.  The action is
 \es{ActionMassiveFermions}{
  S = \int d^3 r \, \sqrt{g(r)}\, \bar \psi \left(i \slashed{D} + i m \right) \psi \,.
 }
The free energy obtained by integrating out $\psi$ is
 \es{FreeFermionsMassive}{
  \tilde {\cal F}_q(m ) = - \tr \log \left(i \slashed{D} + i m \right) \,.
 }
In order to calculate $\tilde {\cal F}_q(m)$ we need to diagonalize the operator $i \slashed{D} + i m$.

The space $H_q$ has an isometry corresponding to shifts in $\tau$ that commutes with the operator $i \slashed{D} + im$, and this isometry allows us to assume that the eigenfunctions of this operator are of the form $e^{i n \tau / q}$ times a spinor on $\HH^2$.  It is not hard to show that for each $n$ there is a continuous spectrum of eigenfunctions parameterized by a positive real number $\lambda$, with eigenvalues
 \es{EvaluesDirac}{
  \pm \sqrt{\lambda^2 + \frac{n^2}{q^2}} + im \,.
 }
The density of states for the continuous parameter $\lambda \geq 0$ is \cite{MR1178146, Bytsenko:1994bc}
 \es{DFermions}{
  \tilde{\cal D}(\lambda) = \frac{\Vol(\HH^2)}{\pi} \lambda \coth(\pi \lambda)  \,.
 }
We require the fermions to be antiperiodic on $S^1$, so $n$ is allowed to take only half odd-integer values.  The free energy is then
 \es{FreeFermions}{
   \tilde {\cal F}_q(m) = -\frac 12 \sum_{n \in \Z + \frac 12} \int_0^\infty d\lambda \, \tilde {\cal D}(\lambda) \log \left[ \lambda^2 + \frac{n^2}{q^2} + m^2 \right] \,.
 }
We can use \eqref{BosonFermionParticular} to do the sum over $n$:
 \es{FreeFermionsAgain}{
  \tilde {\cal F}_q(m) = -\frac 12 \int_0^\infty d\lambda \, \tilde {\cal D}(\lambda)
    \left[ 2 \log \left(1 + e^{-2 \pi q \sqrt{\lambda^2 + m^2} }  \right)
    + 2 \pi q  \sqrt{\lambda^2 + m^2}\right] \,.
 }

We now set $m=0$.  With $\Vol(\HH^2) = -2 \pi$, we have
 \es{FreeFermionsMasslessDiv}{
  \tilde {\cal F}_q \equiv \tilde {\cal F}_q(0) = \int_0^\infty d\lambda \, \lambda \coth(\pi \lambda)
    \left[ 2 \log \left(1 + e^{-2 \pi q \lambda }  \right)
    + 2 \pi q  \lambda \right] \,.
 }
The first term leads to a convergent integral, while the second term needs to be made finite by either zeta-function regularization or by subtracting the flat space free energy density.  Both of these procedures yield the same answer, so we will use the latter:
 \es{SecondTermFerm}{
  2 \pi q  \int_0^\infty d\lambda \, \lambda^2 \coth(\pi \lambda)
   = 2 \pi q  \int_0^\infty d\lambda \, \lambda^2 \left( \coth(\pi \lambda) - 1\right) = q \frac{\zeta(3)}{\pi^2} \,.
 }

We can now write down the free energy of a massless Dirac fermion in terms of a convergent integral:
 \es{FreeFermionsMassless}{
  \tilde {\cal F}_q = 2 \int_0^\infty d\lambda \, \lambda \coth(\pi \lambda)
      \log \left(1 + e^{-2 \pi q \lambda }  \right)
      +q \frac{\zeta(3)}{\pi^2} \,.
 }
Along with the corresponding result for a conformally coupled scalar in eq.~\eqref{ConformalAnswer}, eq.~\eqref{FreeFermionsMassless} is another main result of this paper.

We can check in particular examples that eq.~\eqref{FreeFermionsMassless} reduces to eq.~\eqref{FreeFermionsText} that was derived on the $q$-fold branched covering of $S^3$.  For example, when $q = 1$ we have
 \es{FirstTerm1}{
  \tilde {\cal F}_1 = 2  \int_0^\infty d\lambda \, \lambda \coth(\pi \lambda)
    \log \left(1 + e^{-2 \pi \lambda }  \right) + \frac{\zeta(3)}{\pi^2} = \frac{\log 2}{4} + \frac{3 \zeta(3)}{8 \pi^2} \,,
 }
which is in agreement with eq.~\eqref{FreeFermionsExamples} and \cite{Klebanov:2011gs}.  Also, when $q$ is large the first term in \eqref{FreeFermionsMassless} becomes negligible compared to the second term, in agreement with the large $q$ asymptotics provided in \eqref{FreeLargeq}.

As in the case of bosons, we can check explicitly that the total energy at temperature $T_0 = 1/(2 \pi)$ vanishes:
 \es{FermionEnergy}{
  \tilde E(T_0) = \frac{1}{2 \pi} \frac{d \tilde {\cal F}_q}{dq} \Biggr \rvert_{q=1}
   = -4 \pi  \int_0^\infty d\lambda \,
    \frac{\lambda^2 \coth(\pi \lambda)}{1 + e^{2 \pi \lambda } }+ \frac{\zeta(3)}{\pi^2} = 0 \,.
 }
This result confirms the argument based by conformal mapping to $\R^3$ presented at the end of section~\ref{RENYI}.

\subsection{Results for R\'enyi entropies of a Dirac fermion}

\begin{figure}[htb]
\begin{center}
\leavevmode
\scalebox{1}{\includegraphics{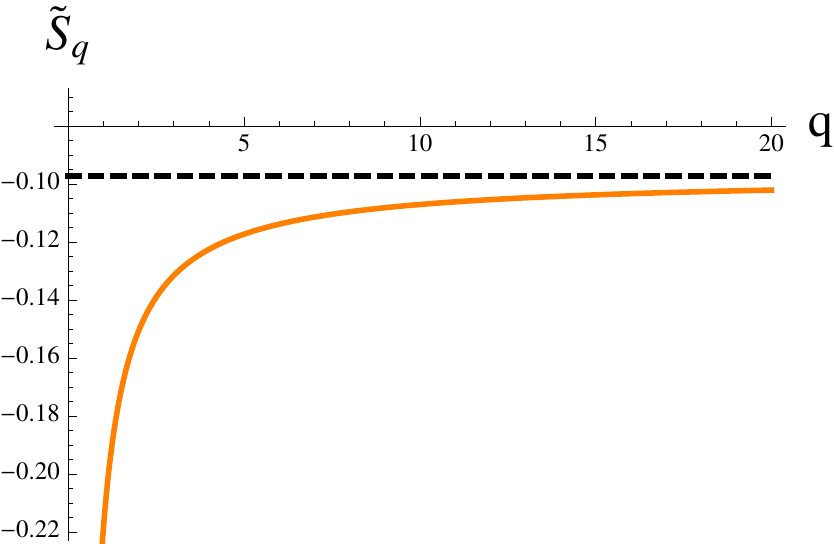}}
\end{center}
\caption{ The R\'enyi entropies $\tilde S_q$ for the massless Dirac field.  Note that the function $\tilde S_q$ is a monotonic function of $q$.  The black dashed line is the asymptotic value $\tilde S_\infty$. }
\label{Sqd}
\end{figure}

Using \eqref{Simple} we can now calculate the R\'enyi entropy of a free massless fermion.  Here are some particular cases:
 \es{RenyiFermions}{
   \tilde S_1 &= -\tilde {\cal F}_1 = - \frac{\log 2}{4} - \frac{3 \zeta(3)}{8 \pi^2} \approx -0.21896 \\
   \tilde S_2 &= \tilde {\cal F}_2 -2 \tilde {\cal F}_1
    = -\frac{5 \log 2}{16} - \frac{21 \zeta(3)}{32 \pi^2} + \frac{G}{2 \pi}  \approx  -0.15076 \,, \\
   \tilde S_3 &= \frac 12 \tilde {\cal F}_3 - \frac 32 \tilde {\cal F}_1
    = -\frac{\log 2}{4} - \frac{13 \zeta(3)}{24 \pi^2} +\frac{\psi_1(1/3)}{6 \sqrt{3} \pi} - \frac{\pi}{9 \sqrt{3}}
      \approx -0.13157 \, \\
   \tilde S_\infty &= -\tilde {\cal F}_1 + \lim_{q \to \infty} \frac{\tilde {\cal F}_q}{q}
    = -\frac{\log 2}{4} + \frac{5 \zeta(3)}{8 \pi^2} \approx -0.09717 \,.
 }
 At large $q$
\es{sqlargeFerm}{
\tilde S_q = \tilde S_\infty \left( 1 + \frac 1 q + O(1/q^2) \right) \, .
}
A plot of $\tilde S_q$ is presented in Fig.~\ref{Sqd}.
We note that, like in the scalar case, $\tilde S_q$ is a monotonic function of $q$.

\section{Massive free fields} \label{MASSIVE}

In this section we study more closely the theories of massive free scalar and Dirac fields on $H_q$.  If the mass of such a theory is denoted by $m$, where $m = 0$ corresponds to the conformal case, then we calculate the functions
\es{fqm}{
f_q (m) \equiv {q {\cal F}_1 (m)- {\cal F}_q(m)\over 1- q}  \,,
}
where ${\cal F}_q (m)$ is the free energy of the theory on $H_q$, with radius set to unity.

While by construction $f_q (0) = S_q$, for generic $m > 0$ the function $f_q (m)$ is not equal to the entanglement entropy across a two-sphere in three flat dimensions.  The mapping of the R\'enyi entropy in $d$-dimensional flat space across an $S^{d-1}$ entangling surface to a calculation on $H_q$ is only valid at conformal fixed points. We will see, however, that in the limit $m\rightarrow \infty$ each $f_q(m)$ falls off to zero exponentially, as is expected of the long range entropy of a massive theory.
The function $f_q (m)$ can therefore be thought of as an interpolating function for the $q$th R\'enyi entropy, between the Gaussian and the trivial conformal fixed points.  We observe for free scalar and Dirac fields that this function is monotonic along the RG flow.

\subsection{Free massive scalars}

\begin{figure}[htb]
\begin{center}
\leavevmode
\scalebox{1}{\includegraphics{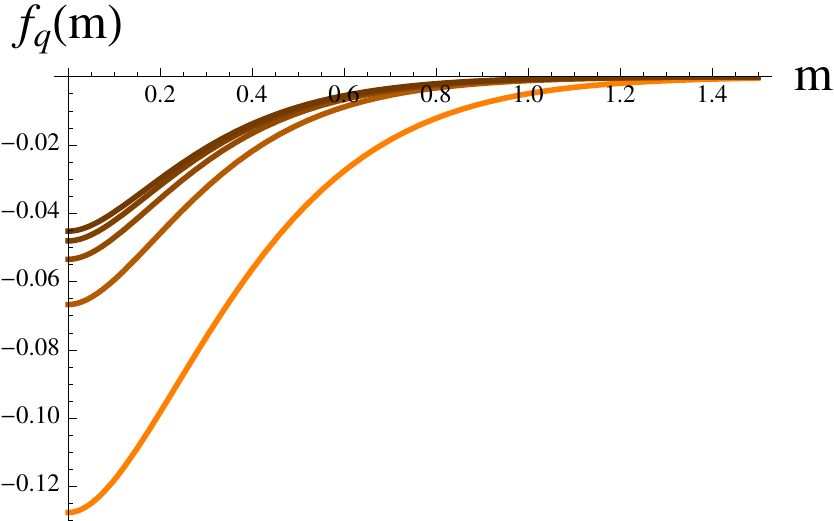}}
\end{center}
\caption{ The R\'enyi entropy interpolating function $f_q(m)$ for a massive complex scalar field, plotted as a function of $m$ for $q = 1,2,3, 4, 5$ (plots are darker for larger $q$). }
\label{Sqmscalar}
\end{figure}

The action for a complex massive free scalar field on $H_q$ is given in equation~\eqref{BosonAction}.  The free energy for this theory on $H_q$ can be found in equation~\eqref{FqOneInt}.  The second term in equation~\eqref{FqOneInt}, corresponding to the temperature-independent vacuum energy, is divergent.  However, this term cancels in~\eqref{fqm}, and thus
$f_q(m)$ is finite.  In figure~\ref{Sqmscalar} we plot the functions $f_q(m)$ for various $q$.
These functions decay exponentially fast in $m$.

\subsection{Free massive Dirac fermions}

\begin{figure}[htb]
\begin{center}
\leavevmode
\scalebox{1}{\includegraphics{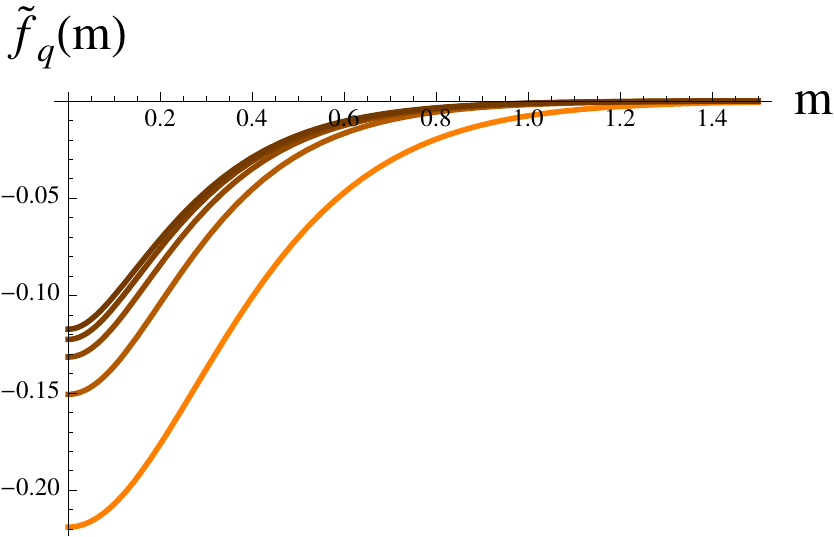}}
\end{center}
\caption{ The R\'enyi entropy interpolating function $\tilde f_q(m)$ for a massive Dirac field, plotted as a function of $m$ for $q = 1,2,3,4,5$ (plots are darker for larger $q$). }
\label{Sqmferm}
\end{figure}

The calculation for free massive Dirac fermions proceeds analogously to that of the scalars.  The action on $H_q$ is given in equation~\eqref{ActionFermions}.  The free energy $\tilde {\cal F}_q (m)$ can be found in equation~\eqref{FreeFermionsAgain}.  Again, while the temperature-independent vacuum energy integral in equation~\eqref{FreeFermionsAgain} is divergent, the function $\tilde f_q (m)$ is finite.  In figure~\ref{Sqmferm} we plot the R\'enyi entropy interpolating function $\tilde f_q (m)$ for various $q$.
The function $\tilde f_q(m)$ decays exponentially fast in $m$ and is stationary at $m = 0$ for all $q$.

\section{Discussion}

In this paper we have presented detailed results for the R\' enyi entropies of $d=2$ free conformal fields across a circle, using conformal mapping methods. Let us note that the approach using the conformal mapping to $S^1 \times \HH^d$ may be easily generalized to $d>2$ \cite{Casini:2010kt,hung} where the entanglement is measured across $S^{d-1}$.
All we need in this approach is the density of states in $\HH^d$, which may be found, for example, in \cite{Bytsenko:1994bc}.\footnote{
For direct numerical calculations of the R\' enyi entropies in $d>3$ see \cite{Braunstein:2011sx}.}

In $d=3$ this approach was used in \cite{Casini:2010kt,Fursaev:2012mp} to calculate the R\' enyi entropies for a free conformal real scalar field \eqref{fourds}. It is not hard to rederive this result using the formulae in our paper.  As in section~\ref{scalarsHyp} the free energy ${\cal F}_q$ is given by equation~\eqref{FqOneInt} with $m=0$ and a density of states ${\cal D}(\lambda)$ appropriate for $\HH^3$ (see for example \cite{camp2, MR1178146, Bytsenko:1994bc}):
\es{D4DS}{
{\cal D} (\lambda) = \frac{ \Vol ( \mathbb{H}^3 ) }{8 \pi^2} \sqrt{\lambda} \,.
}
Note that the density of states on $\mathbb{H}^3$ is identical to that in flat space.  The regulated volume of $\mathbb{H}^3$ is calculated in~\cite{Casini:2010kt} to be
\es{regVol4}{
 \Vol ( \mathbb{H}^3 ) = -2 \pi \log \left(R \over \epsilon\right) .
 }
 After regularization, we then find that the free energy ${ \cal F}_q$ is simply given by
 \es{Fq4s}{
 {\cal F}_q = \frac{1}{360 q^3} \log \left( {R \over \epsilon} \right) \,.
 }
 Upon substitution into (\ref{basicform}) this leads to the R\'enyi entropies (\ref{fourds}).

This approach is also easily applied to a massless Weyl fermion in $d=3$. The free energy $\tilde {\cal F}_q$ of the theory on $S^1 \times \mathbb{H}^3$ is given by equation~\eqref{FreeFermionsAgain} with $m=0$ and density of states \cite{MR1178146, Bytsenko:1994bc}
 \es{D4DW}{
 \tilde {\cal D} (\lambda) = \frac{ \Vol ( \mathbb{H}^3 ) }{\pi^2} \left( \lambda^2 + \frac 1 4 \right) \,.
}
This leads to the free energy
\es{Fq4W}{
\tilde {\cal F}_q = \frac{ 7 + 30 q^2}{1440 q^3} \log \left( { R \over \epsilon}\right) \,.
}
Note that this result agrees with a similar calculation in de Sitter space \cite{Dowker:2010bu}.
Substituting into (\ref{basicform}), we find the R\'enyi entropy
\es{Sq4W}{
\tilde S_q = - \frac{ (1+q)(7+ 37 q^2) }{1440 q^3} \log \left( { R \over \epsilon}\right) \, ,
 }
in agreement with \cite{Fursaev:2012mp}.
The presence of the factor $1+q$ in $d=3$ is guaranteed by the fact that $\tilde {\cal F}_q$ is expanded in odd powers of $q$.

As a check of these results, we note that the universal part of the $d=3$ entanglement entropy satisfies \cite{Solodukhin:2008dh}
\be \label{entanglea}
S_1=- a \log \left( { R \over \epsilon}\right)\,,
\ee
where $a$ is the anomaly coefficient normalized in such a way that $a=1/90$ for a real scalar, and $a=11/180$ for a Weyl fermion.
This general formula follows from the fact that the $q=1$ calculation may be mapped to the partition function on $S^4$ whose logarithmic piece is determined by the Weyl anomaly.
Another consistency check is that in the high temperature limit $q\rightarrow 0$, the curvature of $\mathbb{H}^3$ may be neglected and $S_q$ must be proportional to the standard thermal entropy calculated on $S^1 \times \mathbb{R}^3$. As $q\rightarrow 0$, we find that $\tilde S_q/S_q \rightarrow 7/4$ which is indeed the correct ratio of thermal entropy of a Weyl fermion and a real scalar.

Let us note that both the scalar and the fermion R\'enyi entropies may be expressed as
\es{Renyigen}{
 S_q = - \frac{ (1+q)(B + A q^2) }{360 q^3} \log \left( { R \over \epsilon}\right) \,,
 }
 where $B=F_{\rm therm}/F_{\rm therm}^{\rm scalar}$ and $A=180 a- B$, which follows from a free energy on $S^1 \times \HH^3$ of the form
  \es{Free}{
   {\cal F}_q = \frac{B + (A-B) q^2}{360 q^3} \log \left( { R \over \epsilon}\right) \,.
  }

 It is clear \cite{hung} that in $d=3$ the R\'enyi entropies for $q\neq 1$ are not determined solely by the anomaly coefficient $a$, which is a protected quantity that is not corrected beyond one loop. We have seen that they also contain input from the thermal free energy, which in general receives various perturbative and non-perturbative corrections. Therefore, we do not expect $S_q$ for $q\neq 1$ to satisfy a ``c-theorem.'' In particular, we do not expect them to be constant along lines of fixed points. For example, in the ${\cal N}=4$ SYM theory the R\'enyi entropies should depend on the `t Hooft coupling because the thermal free energy has such dependence \cite{Gubser:1998nz}. The entanglement entropy $S_1$, which is determined by
 $a$ through (\ref{entanglea}), appears to be the only special quantity which is monotonic along RG flows
and is stationary at the fixed points \cite{Cardy:1988cwa,Komargodski:2011vj}. We believe that these statements are valid not only for $d=3$, but for all dimensions $d>1$.
We hope to consider some $d>1$
R\' enyi entropies in more detail in a later publication.

\section*{Acknowledgments}
We thank A. Dymarsky, T.~Grover,
J.~Hung, D.~Jafferis, Z.~Komargodski, J.~Maldacena, R.~Melko, M.~Metlitski, R.~Myers, T.~Nishioka, N.~Seiberg, B.~Swingle, and E.~Witten for helpful discussions.
 The work of IRK was supported in part by the US NSF under Grant No.~PHY-0756966. IRK gratefully acknowledges support from the IBM Einstein Fellowship at the Institute for Advanced Study, and from the John Simon Guggenheim Memorial Fellowship.  SSP was supported by a Pappalardo Fellowship in Physics at MIT and by the U.S. Department of Energy under cooperative research agreement Contract Number DE-FG02-05ER41360\@.  The work of SS was supported by the National Science Foundation under grant DMR-1103860 and by a MURI grant from AFOSR\@.  BRS was supported by the NSF Graduate Research Fellowship Program. BRS thanks the Institute for Advanced Study for hospitality.

\appendix

\section{Degeneracies on $S^3$ and its $q$-fold branched coverings}

\subsection{Setup}

The parameterization \eqref{ParamS3} that yields \eqref{Euclidean} is closely related to the Hopf fibration from the case $q = 1$.  The Hopf fibration is defined as
 \es{Hopf}{
  z_ 1= \cos \frac{\tilde \theta}{2} e^{i (\tilde \phi + \tilde \psi)/2} \,, \qquad
   z_2 = \sin \frac{\tilde \theta}{2} e^{i (-\tilde \phi + \tilde \psi)/2} \,.
 }
So after writing $\theta = \tilde \theta / 2$, $\tau = (\tilde \phi + \tilde \psi)/2$, and $\phi = (-\tilde \phi + \tilde \psi)/2$ the metric becomes
 \es{S3MetricCP1}{
  ds^2 = \frac 14 \left[d\tilde \theta^2 + \sin^2 \tilde \theta\, d \tilde \phi^2 + \left( d\tilde \psi + \cos \tilde \theta\, d\tilde \phi\right)^2 \right] \,,
 }
which exhibits $S^3$ as an $S^1$ fibration over $S^2$.  The ranges of these angles are customarily taken to be $0 \leq \tilde \theta < \pi$, $0 \leq \tilde \phi < 2 \pi$, and $0 \leq \tilde \psi < 4 \pi$.

To work with spinors it is convenient to consider the forms $\sigma_i$ defined as
 \es{sigmaDefs}{
  \sigma_1 + i \sigma_2 = e^{i \phi + i\tau} \left[ i d \theta + \sin \theta \cos \theta (d\tau - d \phi) \right] \,, \qquad
   \sigma_3 =\cos^2 \theta\, d \tau + \sin^2 \theta d\phi \,.
 }
These $\sigma_i$ are the left-invariant one forms on $S^3$, and away from $\theta = 0, \pi/2$ they are well-defined and non-vanishing. They satisfy
 \es{sigmaRel}{
  d\sigma_i = \epsilon_{ijk} \sigma_j \wedge \sigma_k \,,
 }
and, in addition, the metric on $S^3$ can be written as
 \es{S3MetricSigma}{
  ds^2 = \sigma_1^2 + \sigma_2^2 + \sigma_3^2  \,.
 }
We will henceforth use the left-invariant forms $\sigma_i$ as a dreibein.

On the $q$-fold branched covering of $S^3$ \eqref{Euclidean} with $q>1$ one can still use the forms $\sigma_i$ given in \eqref{sigmaDefs} as a dreibein.  The only difference is that now the range of $\tau$ is $[0, 2 \pi q)$, with $q$ taken to be a positive integer.

\subsection{Degeneracies for fermions}
 \label{DEGFERMIONS}

To study fermions, let us take the gamma matrices in 3d to be equal to the Pauli matrices:
 \es{Pauli}{
  \gamma_1 = \begin{pmatrix}
   0 & 1 \\
   1 & 0
   \end{pmatrix} \,, \qquad
    \gamma_2 = \begin{pmatrix}
     0 & -i \\
     i & 0
    \end{pmatrix} \,, \qquad
     \gamma_3 = \begin{pmatrix}
      1 & 0 \\
      0 & -1
     \end{pmatrix} \,.
 }
That the spinor $\psi$ is an eigenspinor of the Dirac operator with eigenvalue $\lambda$ means that
 \es{psiEigen}{
  i \sigma^\mu_a \gamma^a \left( \partial_\mu \psi + \frac 14 \omega_\mu^{ab} \gamma_{ab} \psi \right) = \lambda \psi \,.
 }

For solving \eqref{psiEigen} we make the ansatz
 \es{FermionAnsatz}{
  \psi = e^{i m \tau + i n \phi} \begin{pmatrix}
   e^{-(\phi + \tau)/2} f_+(\theta) \\
   e^{(\phi + \tau)/2} f_-(\theta)
  \end{pmatrix} \,.
 }
The solution that is square-integrable at $\theta = \pi/2$ can be written as
 \es{FermionSoln}{
  f_-(\theta) = \frac 12 \left[ (2m - 1) \tan \theta - (2n-1) \cot\theta \right] f(\theta) + f'(\theta) \\
  f_+(\theta) = \left( m + n - \lambda - \frac 12 \right) f(\theta) \,,
 }
where
 \es{freg}{
  f(\theta) = \begin{cases}
   (\cos \theta)^{m - \frac 12} (\sin \theta)^{-n + \frac 12} {}_2 F_1\left(\frac{m - n - \lambda + \frac 12}{2},
    \frac{m - n + \lambda + \frac 32}{2}; m + \frac 12; \cos^2 \theta \right) & \text{if } m \geq 0\\
    (\cos \theta)^{\frac 12 - m} (\sin \theta)^{-n + \frac 12} {}_2 F_1\left(\frac{-m - n - \lambda + \frac 32}{2},
    \frac{-m - n + \lambda + \frac 52}{2}; -m + \frac 32; \cos^2 \theta \right) & \text{if }m <0
  \end{cases} \,.
 }
Imposing regularity at $\theta = 0$ as well restricts the allowed values of $\lambda$ in terms of $n$ and $m$.  In particular, there are several infinite series of eigenmodes:  two with positive $\lambda$,
 \es{CaseI}{
  \lambda &= m + n + \frac 32 + 2a \,, \qquad m \geq 0\,, \qquad n \geq -\frac 12-a \,, \qquad a \in \N \,, \\
  \lambda &= -m + n + \frac 12 + 2a \,, \qquad m <0 \,, \qquad n \geq \frac 12-a \,, \qquad a \in \N \,,
 }
and two with negative $\lambda$:
 \es{CaseII}{
  \lambda &= -\left(m + n + \frac 12 + 2a \right) \,, \qquad m \geq  0\,, \qquad n \geq \frac 12-a \,, \qquad a \in \N \,, \\
  \lambda &= -\left( -m + n + \frac 32 + 2a\right) \,, \qquad m <0 \,, \qquad n \geq -\frac 12-a \,, \qquad a \in \N \,.
 }

To completely solve the eigenvalue problem \eqref{psiEigen} on the $q$-fold branched covering of $S^3$, we need to decide on the allowed values of $n$ and $m$ by requiring the spinor \eqref{FermionAnsatz} to be well-defined.  Under either $\tau \to \tau + 2 \pi q$ or $\phi \to \phi + 2 \pi$ we must have $\psi \to -\psi$, which implies $e^{2 \pi i n} = e^{2 \pi i m q} = -1$.  Consequently, $n \in \Z + \frac 12$ and $m \in \frac 1q \left( \Z + \frac 12 \right)$.
It can be proven by induction that the eigenvalues of the Dirac operator then are those given in eq.~\eqref{DiracEvalues} with multiplicity $(k+1)(k+2)$ for each choice of sign.

\subsection{Vector fields}
\label{VECTOR}

We can also find a complete basis of vector fields, or equivalently one-forms, on $S^3$.  To do so, we find the eigenfunctions of the vector Laplacian
 \es{VectorLap}{
  \Delta = *d*d + d*d* \,,
 }
and these eigenfunctions will be our basis.  Hodge decomposition states that any one-form $A$ can be written as
 \es{ClosedForms}{
  A = B + d \phi + H \,, \qquad d*B = 0 \,, \qquad \Delta H = 0
 }
for some $\phi$, $B$, and $H$.  Since $H^1(S^3) = 0$, there are no non-zero harmonic one-forms $H$ on $S^3$.
So in order to find a basis of forms $A$, it is enough to find a basis of exact one-forms $d\phi$ and a basis of co-closed one-forms $B$.

The basis of exact one-forms is given by derivatives of the spherical harmonics.  Indeed, if $Y$ is a spherical harmonic satisfying $\nabla^2 Y = -*d*dY = -n(n+2) Y$, then $dY$ is of course exact and is an eigenfunction of $\Delta$:
 \es{YIsEigenfunction}{
  \Delta dY = d*d*dY = n(n+2) dY \,.
 }

Diagonalizing $\Delta$ in the subspace of co-closed one-forms is easier on a three-dimensional manifold than on a manifold of arbitrary dimension because only in three dimensions does the operator $*d$ map one-forms to one-forms.  In fact $*d$ is a self-adjoint operator on $\Omega^1(S^3)$.  Suppose we find a co-closed $B$ that is an eigenfunction of $*d$ with eigenvalue $\lambda$:
 \es{EqsToSolve}{
  d*B = 0 \,, \qquad *dB = \lambda B \,,
 }
Then $B$ is also an eigenfunction of $\Delta$:
 \es{BIsEigenfunction}{
  \Delta B = (*d)^2B = \lambda^2 B\,.
 }

In order to solve the equations \eqref{EqsToSolve} we make the ansatz
 \es{BruteForceAnsatz}{
  B = e^{i m \tau + i n \phi} \left[e^{-i (\tau + \phi)} b_+(\theta) \sigma_+ + b_0(\theta) \sigma_3  + e^{i (\tau + \phi)} b_-(\theta) \sigma_- \right] \,.
 }
For any given $\lambda$, the solution that is regular at $\theta = \pi/2$ is:
 \es{RegularSoln}{
  b_\pm (\theta) &=  (m+n \mp\lambda) \left[\mp\frac 12  f'(\theta) + \frac 12 f(\theta) \left(-n \cot \theta + m \tan \theta \right) \right] \,, \\
  b_0(\theta) &= (m+n+\lambda)(m+n-\lambda) f(\theta) \,,
 }
where
 \es{Gotf}{
  f(\theta) = (\cos \theta)^{\abs{m}} (\sin \theta)^n
    {}_2F_1\left(\frac{\abs{m} + n + \lambda}{2}, \frac{2 + \abs{m} + n - \lambda}{2}; 1 + \abs{m}; \cos^2 \theta \right) \,.
 }
Requiring that \eqref{RegularSoln} is regular at $\theta = 0$ yields four towers of solutions:  two with positive $\lambda$:
 \es{GotLambda}{
  \lambda &= m - n + 2 +2a\,, \qquad m \geq 0 \,, \qquad n \leq a+1 \\
  \lambda &= 2- m + n + 2a \,, \qquad m \leq 0 \,, \qquad n \geq a-1 \,,
 }
where $a \in \N$, and two similar towers with negative $\lambda$.

Counting carefully we find that the eigenvalues of $*d$ in the space of co-closed one-forms are of the form $\pm n$, with degeneracies
 \es{DegeneraciesVector}{
  g_n = \begin{cases}
   n^2 - 1 & \text{if } n \in \Z \,, \quad n \geq 2 \,, \\
   m(m+1) & \text{if } n = m + \frac pq \,, \quad m \in \Z \,, \quad m \geq 1 \,,
  \end{cases}
 }
for every integer $p$ that satisfies $1 \leq p < q$.

\section{Useful mathematical formulae}
\label{MATH}

In this section we present some useful mathematical formulae.  We begin with zeta function identities.
For $0 < a \leq 1$ we have the identity
 \es{ZetaIdentity}{
  \zeta(z, a) = \frac{2 \Gamma(1 - z)}{(2 \pi)^{1-z}} \left[\sin \frac{z \pi}{2} \sum_{n=1}^\infty \frac{\cos 2 \pi a n}{n^{1-z}}
   + \cos \frac{z \pi}{2} \sum_{n=1}^\infty \frac{\sin 2 \pi a n}{n^{1-z}} \right] \,.
 }
Taking derivatives at $z=0, -1, -2$ gives
 \es{DerZetaIdentity}{
  \zeta'(-2, a) &= - \frac{1}{4 \pi^2} \sum_{n=1}^\infty \frac{\cos 2 \pi a n}{n^3}
   - \frac{1}{4 \pi^3} \sum_{n=1}^\infty \frac{(2 \log (2 \pi n) + 2 \gamma - 3) \sin 2 \pi a n}{n^3} \,, \\
  \zeta'(-1, a) &= \frac{1}{4 \pi} \sum_{n=1}^\infty \frac{\sin 2 \pi q n}{n^2}
   - \frac{1}{2 \pi^2} \sum_{n=1}^\infty \frac{(\log (2 \pi n) + \gamma - 1) \cos 2 \pi a n}{n^2} \,, \\
  \zeta'(0, a) &= \frac{1}{2} \sum_{n=1}^\infty \frac{\cos 2 \pi a n}{n}
   + \frac{1}{\pi} \sum_{n=1}^\infty \frac{(\log (2 \pi n) + \gamma) \sin 2 \pi a n}{n} \,.
 }

Two other useful identities are the regularized sums
 \es{BosonFermionParticular}{
   \sum_{n \in \Z} \log \left( \frac{n^2}{q^2} + a^2 \right)
   &=  2 \log \left[2 \sinh (\pi q \abs{a}) \right] \,, \\
   \sum_{n \in \Z + \frac 12} \log \left(\frac{n^2}{q^2} + a^2 \right)
   &=  2 \log \left[2 \cosh (\pi q \abs{a}) \right] \,.
 }
These sums follow from the more general formula
 \es{SumBF}{
   \sum_{n \in \Z} \log \left( \frac{(n + \alpha)^2}{q^2} + a^2 \right)
   &=  \log \left[2 \cosh (2 \pi q \abs{a}) - 2 \cos (2 \pi \alpha)  \right] \,.
 }
This relation in turn follows from the Poisson summation formula
 \es{Poisson}{
 \frac{1}{ 2 \pi q} \sum_{n \in \Z} \hat f \left( \frac{n + \alpha}{q} \right) =\sum_{k \in \Z} e^{-i 2 \pi k \alpha} f(2 \pi q k)  \,
 }
applied to
 \es{fhat}{
  \hat f(\omega)  = \log \left( \omega^2 + a^2 \right) \,.
 }
For $t \neq 0$ one can simply calculate the inverse Fourier transform of $\hat f$:
 \es{fnonzero}{
  f(t) = \int_{-\infty}^\infty \frac{d\omega}{2 \pi} e^{-i \omega t} \log \left(\omega^2 +a^2 \right) = - \frac{e^{-\abs{a} \abs{t}}}{\abs{t}}  \,.
 }
The case $t=0$ requires special care because the expression for $f(0)$ is divergent and requires regularization:
 \es{fzero}{
  f(0) = \int_{-\infty}^\infty \frac{d\omega}{2 \pi} \log \left(\omega^2 +a^2 \right)
   = -\frac{d}{ds} \int \frac{d\omega}{2 \pi} \frac{1}{\left(\omega^2 +a^2 \right)^s} \Biggr\rvert_{s=0}
    = \abs{a} \,.
 }
Using \eqref{fhat}--\eqref{fzero} one can show that eq.~\eqref{Poisson} reduces to eq.~\eqref{SumBF}.

\section{Chern-Simons theory on $C_q$} \label{cssection}
\label{app:cs}

In this section we evaluate the free energy ${\cal F}_{\text{CS}}$ of $U(1)$ Chern-Simons (CS) theory on $C_q$ and reproduce the known result that ${\cal F}_{\text{CS}}$ is independent of $q$.   The partition function of CS theory is
 \es{PartFunction}{
  Z_\text{CS} = \frac{1}{2 \pi \sqrt{\Vol(C_q)}} \int DA e^{-S_\text{CS} - S_\text{ghost}} \,,
 }
where $S_\text{CS}$ is the Chern-Simons action
\es{CSaction}{
S_\text{CS} = \frac{ik}{4\pi} \int A \wedge d A \,,
}
and $S_\text{ghost}$ is an action involving ghosts required for proper gauge fixing.  We use the gauge $d * A = 0$, in which case the ghost action is
 \es{GhostAction}{
  S_\text{ghost} = \int d^3 x\, \sqrt{g} \left[ \bar c \nabla^2 c + b \nabla^\mu A_\mu \right] \,,
 }
where $c$ is a fermionic ghost and $b$ is a bosonic ghost.  The factor of $1/\sqrt{\Vol(C_q)}$ in the partition function is important in order for the answer to be independent of the metric on $C_q$.  The factor of $1/(2 \pi)$ comes from the volume of the gauge group, which one customarily divides by.

The partition function \eqref{PartFunction} yields the free energy
 \es{FreeCSAgain}{
  {\cal F}^\text{CS}_q = {\cal F}^{\text{vec}}_q + {\cal F}^\text{ghost}_q + \log (2 \pi) + \frac 12 \log (q \Vol(S^3)) \,,
 }
where we used the fact that $\Vol(C_q) = q \Vol(S_3)$, and denoted by ${\cal F}^{\text{vec}}_q$ the contribution from the Chern-Simons action \eqref{CSaction} in the gauge $d*A = 0$, and by ${\cal F}^\text{ghost}_q$ the contribution from the ghosts.  It is straightforward to show that ${\cal F}^\text{ghost}_q$ equals minus the free energy of a massless complex scalar on $C_q$, while ${\cal F}^\text{vec}_q$ can be computed from the eigenvalues and degeneracies in \eqref{DegeneraciesVector}.  When $q=1$ we have
 \es{FGhost}{
  {\cal F}^{\text{vec}}_{1} &= \sum_{n=2}^\infty (n^2 - 1) \log \frac{kn}{4 \pi^2}  = \frac 12 \log \frac{k}{8 \pi^3} + \frac{\zeta(3)}{4 \pi}\,, \\
  {\cal F}^\text{ghost}_{1} &= \frac 12 \sum_{n=1}^\infty (n+1)^2 \log n(n+2) = \frac 12 \log \pi + \frac{\zeta(3)}{4 \pi} \,,
 }
which implies ${\cal F}^\text{CS}_1 = \frac 12 \log k$.  For the $q$th branched covering of $S^3$ we have:
 \es{FCSCovering}{
  {\cal F}^\text{CS}_q &= {\cal F}^\text{CS}_1 + \frac{\log q}{2} +  \sum_{p=1}^{q-1} \sum_{m=1}^\infty m(m+1) \left[ \log \frac{k \left(m+\frac pq\right)}{4 \pi^2}
   - \frac 12 \log \left(m-1 + \frac pq\right) \left(m+1 + \frac pq\right)\right]  \\
   &= {\cal F}^\text{CS}_1 + \frac{\log q}{2} + \sum_{p=1}^{q-1} \zeta'(0, p/q) = {\cal F}^\text{CS}_1 \,.
 }

The fact that ${\cal F}^\text{CS}_q = {\cal F}^\text{CS}_1$ for all $q$ can be understood from the fact that the topology of the branched covering $C_q$ is the same as that of $S^3$.  The Chern-Simons partition function is a topological invariant, so it should indeed be independent of $q$.  From eq.~\eqref{basicform} it follows that all the R\'enyi entropies are equal:
 \es{RenyiCS}{
  S_q^\text{CS} = S_1^\text{CS} = - \frac 12 \log k \,.
 }

\bibliographystyle{ssg}
\bibliography{Renyipaper}

\end{document}